\documentclass[fleqn,usenatbib,useAMS]{mnras}
 \usepackage{graphicx}
 \usepackage{amsmath}
 \usepackage{amssymb}
 \usepackage[T1]{fontenc}
 \usepackage{ae,aecompl}
 \usepackage{txfonts}
 \usepackage{enumerate}
 \title[{\it TESS} TNO candidates under scrutiny]
       {Distant trans-Neptunian object candidates from NASA's \textit{\textbf{TESS}} 
        mission scrutinized: fainter than predicted or false positives?\thanks{Based 
        on service observations (proposals SW2021a24 and SW2021a10) made with the 
        4.2-m {\it William Herschel} Telescope ({\it WHT}), operated by the 
        {\it Isaac Newton} Group at the Observatorio del Roque de los Muchachos, La 
        Palma, Spain, of the Instituto de Astrof\'{\i}sica de Canarias.}
       }

 \author[C. de la Fuente Marcos et al.]
        {C.~de~la~Fuente~Marcos,$^{1}$\thanks{E-mail: nbplanet@ucm.es}
         R. de la Fuente Marcos,$^{2}$ 
         O.~Vaduvescu$^{3, 4, 5}$
         and
         M.~St{\u{a}}nescu$^{6}$ \\
         $^1$Universidad Complutense de Madrid,
             Ciudad Universitaria, E-28040 Madrid, Spain \\
         $^2$AEGORA Research Group,
             Facultad de Ciencias Matem\'aticas,
             Universidad Complutense de Madrid,
             Ciudad Universitaria, E-28040 Madrid, Spain \\
         $^3$Isaac Newton Group (ING),
             Apt. de correos 321, E-38700, Santa Cruz de La Palma, Canary Islands, Spain \\
         $^4$Instituto de Astrof\'{\i}sica de Canarias (IAC),
             C/ V\'{\i}a L\'actea s/n, E-38205 La Laguna, Tenerife, Spain \\
         $^5$University of Craiova,
             Str. A. I. Cuza nr. 13, 200585, Craiova, Romania \\
         $^6$Bucharest Astroclub,
             Str. Cutitul de Argint 5, sector 4, 052034,Bucharest, Romania}
 \date{Accepted 2022 April 5. 
       Received 2022 March 27; 
       in original form 2022 March 6}
 \pubyear{2022}
 
\begin{document}
  \label{firstpage}
  \pagerange{\pageref{firstpage}--\pageref{lastpage}}
  \maketitle
%
%
  \begin{abstract}
     NASA's {\it Transiting Exoplanet Survey Satellite} ({\it TESS}) is 
     performing a homogeneous survey of the sky from space in search of 
     transiting exoplanets. The collected data are also being used for 
     detecting passing Solar system objects, including 17 new outer Solar 
     system body candidates located at geocentric distances in the range 
     80--200~au, that need follow-up observations with ground-based telescope 
     resources for confirmation. Here, we present results of a 
     proof-of-concept mini-survey aimed at recovering two of these candidates 
     that was carried out with the 4.2-m {\it William Herschel} Telescope and 
     a QHY600L~CMOS camera mounted at its prime focus. For each candidate 
     attempted, we surveyed a square of over 1{\degr}$\times$1{\degr} around 
     its expected coordinates in Sloan $r'$. The same patch of sky was 
     revisited in five consecutive or nearly consecutive nights, reaching 
     S/N$>$4 at $r'$$<$23~mag. We focused on the areas of sky around the 
     circumpolar {\it TESS} candidates located at 
     (07$^{\rm h}$:00$^{\rm m}$:15$^{\rm s}$, 
     +86{\degr}:55{\arcmin}:19{\arcsec}), 202.8~au from Earth, and 
     (06$^{\rm h}$:39$^{\rm m}$:47$^{\rm s}$, 
     +83{\degr}:43{\arcmin}:54{\arcsec}) at 162.1~au, but we could not recover 
     either of them at $r'$$\leq$23~mag. Based on the detailed analysis of the 
     acquired images, we confirm that either both candidates are much fainter 
     than predicted or that they are false positives.
  \end{abstract}


  \section{Introduction\label{intro}}
     Our degree of understanding of the structure of the Solar system beyond the trans-Neptunian or Kuiper belt and out into the 
     Oort cloud remains very limited. The discoveries of 2018~VG$_{18}$, found at a heliocentric distance of about 123~au with 
     unfiltered magnitude 24.6 \citep{2018MPEC....Y...14S}, and 2018~AG$_{37}$, found at 132~au with $G$=25.3~mag 
     \citep{2021MPEC....C..187S}, represent the opening of a new window into this remote region, the Solar system beyond 100~au 
     from the Sun. Although just two objects have been detected with current distances $>$100~au, they are both following highly 
     eccentric orbits and are arguably similar (in terms of their orbital properties) to several other confirmed Centaurs and 
     scattered disc objects.

     NASA's {\it Transiting Exoplanet Survey Satellite} 
     ({\it TESS})\footnote{\href{https://heasarc.gsfc.nasa.gov/docs/tess/primary.html}
     {https://heasarc.gsfc.nasa.gov/docs/tess/primary.html}} is performing a homogeneous survey of the sky from space in search
     of transiting exoplanets \citep{2015JATIS...1a4003R} ---since 2018 it has observed approximately 80 per cent of the sky--- 
     but it is also capable of detecting passing Solar system small bodies (see e.g. \citealt{2018PASP..130k4503P,
     2019RNAAS...3..160H,2019ApJS..245...29M,2019RNAAS...3..172P,2020ApJS..247...26P,2021PASP..133a4503W}). By shift-stacking
     {\it TESS} images, \citet{2020PSJ.....1...81R} were able to recover three previously known outer Solar system bodies: 90377 
     Sedna (2003~VB$_{12}$) at about 84~au from the Sun with $V$=20.64~mag, 2015~BP$_{519}$ at about 54~au with $V$=21.81~mag, and
     523622 (2007~TG$_{422}$) at about 37~au with $V$=22.32~mag. \citet{2020PSJ.....1...81R} then applied a blind search
     shift-stacking algorithm to {\it TESS} sectors 18 and 19, and singled out 17 new outer Solar system body candidates, which 
     need to be followed up with ground-based observations for confirmation. Their technique combines (stacks) a series of images, 
     acquired at different times, to turn an undetected source into a brighter, single point-like occurrence. The stacking is done 
     along different paths (shifts), assuming that the optimal one leads to the highest S/N for the newly detected source. 

     The {\it TESS} candidate objects are located at geocentric distances in the range 80--200~au; the list of candidates in 
     table~2 of \citet{2020PSJ.....1...81R} includes eight objects located farther than 100~au from the Sun (referred to epochs 
     2458838.92~JD or 10:4:48.00 UT on December~21, 2019 and 2458810.25~JD or 18:0:0.00 UT on November~22, 2019) that, if 
     confirmed, would join 2018~VG$_{18}$ and 2018~AG$_{37}$ as the farthest Solar system objects ever observed. Confirming the 
     existence of any of these eight distant candidates can significantly increase our knowledge of the outer Solar system as 
     proper orbit determinations (the published data give only their instantaneous positions over two years ago) may enable their 
     eventual physical and dynamical characterization (see e.g. \citealt{2017MNRAS.467L..66D}). Furthermore, the study of 
     additional distant trans-Neptunian objects (TNOs) may help in confirming or rejecting the presence of putative planetary 
     bodies beyond the trans-Neptunian belt (see e.g. \citealt{2014MNRAS.443L..59D,2014Natur.507..471T,2016AJ....151...22B}).

     In this Letter, we present results of a proof-of-concept mini-survey carried out with the 4.2-m {\it William Herschel}
     Telescope ({\it WHT})\footnote{\href{https://www.ing.iac.es/astronomy/telescopes/wht/}
     {https://www.ing.iac.es/astronomy/telescopes/wht/}} and a QHY600L CMOS camera mounted temporarily on-axis at the prime focus
     of the {\it WHT} (PF-QHY)\footnote{\href{https://www.ing.iac.es//Astronomy/instruments/pf-qhy/}
     {https://www.ing.iac.es//Astronomy/instruments/pf-qhy/}}. The survey was aimed at recovering two of the most distant new 
     outer Solar system body candidates found in data from NASA's {\it TESS} mission and discussed by \citet{2020PSJ.....1...81R}. 
     In Section~2, we describe the observations acquired and in Section~3, we present the results of our analyses. A discussion is 
     presented in Section~4. Finally, our conclusions are summarized in Section~5.
%
%
     \begin{figure*}
        \centering
        \includegraphics[width=0.49\linewidth, trim= 6 0 5 0, clip]{candidate_C_2.pdf}
        \includegraphics[width=0.49\linewidth, trim= 6 0 5 0, clip]{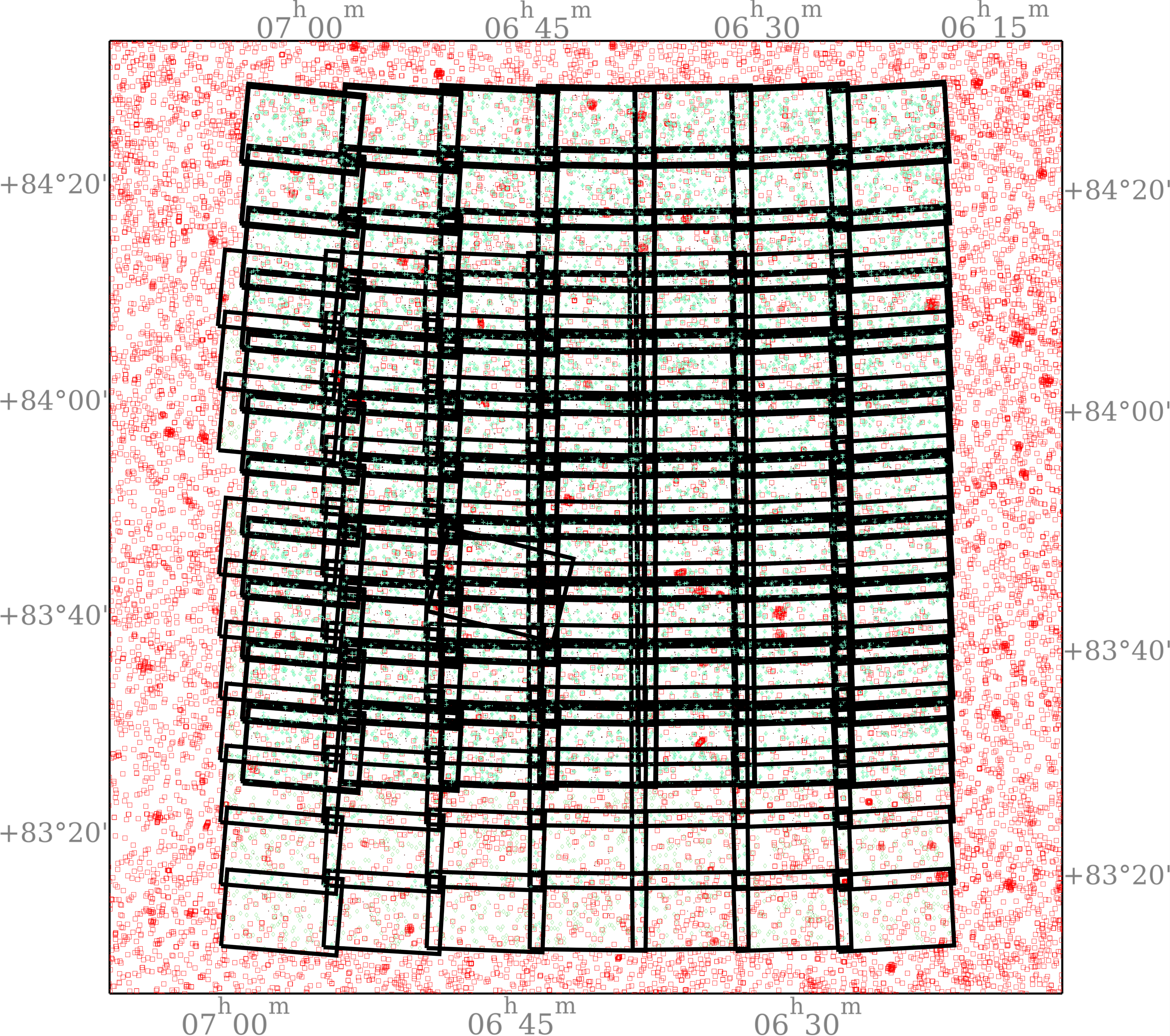}
        \caption{Fields observed by our survey. In both panels, North is up, East to the left. All the pointings 
                 (2 fields$\times$77 pointings per field$\times$5 nights per field) have been plotted. The left-hand side panel 
                 shows the field in Cepheus and centred towards ($\alpha$, $\delta$) = (105{\fdg}0573, +86{\fdg}9216), while the 
                 right-hand panel displays the field in Camelopardalis and centred towards (99{\fdg}9219, +83{\fdg}7321). The 
                 single, oddly rotated pointing was due to a short-lived technical issue and it does not affect our conclusions. 
                 Pointings for July~8 appear somewhat shifted due to another, different technical issue but this does not have an 
                 impact on our conclusions because data from the fifth night were only used as a redundant consistency check.}
        \label{fields}
     \end{figure*}
%
%

  \section{Observations}
     We obtained Sloan $r'$ CMOS images of two fields --- the first one in Cepheus and the second one in Camelopardalis (see
     Fig.~\ref{fields}) --- as squares of over 1{\degr}$\times$1{\degr} with {\it WHT} on July 2, 3, 4, 6, and 8, 2021. The QHY 
     camera (model QHY600L)\footnote{\href{https://www.qhyccd.com/qhy600m-c/}{https://www.qhyccd.com/qhy600m-c/}} used in this
     proof-of-concept mini-survey is based on a back-illuminated CMOS detector
     (Sony IMX455)\footnote{\href{https://www.sony-semicon.co.jp/products/common/pdf/IMX455AQK-K\_Flyer.pdf}
     {https://www.sony-semicon.co.jp/products/common/pdf/IMX455AQK-K\_Flyer.pdf}}, with 9576$\times$6388 3.8-$\mu$m pixels, giving 
     a field of view of 10{\farcm}7$\times$7{\farcm}1 with a scale of 0{\farcs}067~pixel$^{-1}$ at the prime focus of the 
     {\it WHT}; with the recommended binning of 4$\times$4 that was used in our study, the scale became 0{\farcs}267~pixel$^{-1}$. 
     The camera was tilted by 90{\degr} so the longest dimension of the CMOS chip would be along the right ascension axis, 
     optimizing the scanning process in terms of time used on-sky.
%
%
     \begin{table}
        \centering
        \fontsize{8}{12pt}\selectfont
        \tabcolsep 0.15truecm
        \caption{\label{observations}Observations from {\it WHT}. Information includes date, Julian date, time used on-sky, 
                 {\it WHT} seeing range (Differential Image Motion Monitor, DIMM, values), and dark time window during the night. 
                 The images were all acquired during dark time.}
        \begin{tabular}{lcccc}
           \hline
           Date   & Julian date & UT-range       & seeing     & Dark time     \\
           (2021) &             & (h:m) UT       & (\arcsec)  & (h:m) UT      \\
           \hline
           July 2 & 2459398     & 22:40 -- 00:45 & 0.8 -- 1.3 & 21:48 -- 1:43 \\
           July 3 & 2459399     & 21:45 -- 23:55 & 0.6 -- 1.1 & 21:48 -- 2:13 \\
           July 4 & 2459400     & 21:38 -- 00:24 & 0.6 -- 1.2 & 21:47 -- 4:45 \\
           July 6 & 2459402     & 22:43 -- 00:50 & 0.6 -- 0.9 & 21:47 -- 4:46 \\
           July 8 & 2459404     & 02:30 -- 04:30 & 1.0 -- 2.0 & 21:47 -- 4:47 \\
           \hline
        \end{tabular}
     \end{table}
%
%

     Each square patch of sky was densely sampled with 77 pointings --- each one of 10{\farcm}7$\times$7{\farcm}1 and with
     sufficient overlapping between neighbouring pointings (see Fig.~\ref{fields}). The 77 pointings of both square patches were 
     repeated five nights (see Table~\ref{observations}). The pointings were carried out using a \textsc{Python} 
     \citep{van1995python} script, scanning the patch of sky of slightly over one square degree after reading the coordinates of 
     the centre of the field as input. Each individual pointing had an exposure time of 30~s with sidereal tracking. This choice 
     should produce detections with S/N$>$4 at Sloan $r'$$<$23~mag in good seeing ($\leq$1{\farcs}2, see the discussion in 
     Section~4). Due to the circumpolar nature of the target fields, they were observed at low elevations in the range 
     20{\degr}--30{\degr} --- airmasses in the range 2.21--2.27 for the field in Cepheus and 2.52--2.61 for the one in 
     Camelopardalis. The target candidates move at a rate of the order of 1{\arcsec}~h$^{-1}$; therefore, if a target field is 
     revisited within 24~h, it should be relatively straightforward to identify the moving objects (if they are present in the 
     imaged fields) by visually blinking the images for example. In our case, detections were flagged via software and 
     subsequently validated or rejected after human analysis.

     The data were processed by the automatic pipelines of the \textsc{Umbrella} suite \citep{2021A&C....3500453S}, with
     human-validated detections. The processing of the raw frames into detection-ready images was carried out with the
     \textsc{Umbrella IPP} (Image pre-Processing Pipeline) that is a new component, not described in the original 
     \textsc{Umbrella} paper. This new component uses dark and flat processing built on top of \textsc{Umbrella2} 
     \citep{2020ascl.soft08006S} and plate solving and field correction through \textsc{AstrOmatic} \citep{1996A&AS..117..393B,
     2002ASPC..281..228B,2006ASPC..351..112B}. Dark images were used to remove any hot pixels. The detection pipeline applied was 
     the reference blink pipeline (with tweaks and improvements specifically designed for this survey) called through the 
     \textsc{Webrella} component, a web-based front-end application that allowed the entire collaboration to have quick access to 
     the pipeline output. All the processing was performed on the server of the EUROpean Near Earth Asteroids Research 
     (EURONEAR)\footnote{\href{http://www.euronear.org/}{http://www.euronear.org/}} project at the University of Craiova. In case 
     of recovering a putative candidate, its nature could be confirmed statistically using e.g. the software discussed by 
     \citet{2000AJ....120.3323B}.

  \section{Results}
     The two fields observed correspond to those of candidates~9 and 11 in table~2 of \citet{2020PSJ.....1...81R}. Candidate~9 is
     the farthest one in their list, it is located at a geocentric distance of 202.8~au towards ($\alpha$, $\delta$) =
     (105{\fdg}0573, +86{\fdg}9216) in Cepheus, and it has $V$=21.77~mag. Candidate~11 is the third farthest in
     \citet{2020PSJ.....1...81R} with a geocentric distance of 162.1~au and observed projected towards (99{\fdg}9219,
     +83{\fdg}7321) in Camelopardalis with $V$=21.92~mag. From the data in \citet{2020PSJ.....1...81R}, the ecliptic latitude of
     candidate~9 is 63{\fdg}6 and that of candidate~11 is 60{\fdg}4; being both so distant, their orbital inclinations, $i$, must
     be significant and this fact adds another intriguing property to these candidates. The confirmed extreme TNO (ETNO)
     2015~BP$_{519}$ --- also detected by \citet{2020PSJ.....1...81R} --- is considered an outlier with $i$=54{\degr}
     \citep{2018AJ....156...81B,2018RNAAS...2..167D} and it was observed at an ecliptic latitude close to $-$54{\degr} when
     discovered. In addition to being outliers in terms of orbital inclination, the candidates (if real) could be outliers in 
     terms of size (assuming typical values for the albedo) as they would both be distant and comparatively bright within the 
     context of the known TNO populations.

     In addition to the \textsc{Umbrella2}-based automated detection pipeline, we used the
     \textsc{Astrometrica}\footnote{\href{http://www.astrometrica.at/}{http://www.astrometrica.at/}} software
     \citep{2012ascl.soft03012R} in automated and visual blinking mode, as well as pure visual blinking with
     DS9\footnote{\href{https://sites.google.com/cfa.harvard.edu/saoimageds9}
     {https://sites.google.com/cfa.harvard.edu/saoimageds9}} \citep{2003ASPC..295..489J}, to search for moving objects in both
     fields. Automated and visual analysis of the data collected did not produce any moving object detections. Querying the
     \textsc{MPChecker}\footnote{\href{https://minorplanetcenter.net/cgi-bin/checkmp.cgi}
     {https://minorplanetcenter.net/cgi-bin/checkmp.cgi}} tool of the Minor Planet Center (MPC,
     \citealt{2016IAUS..318..265R,2019AAS...23324503H})\footnote{\href{https://minorplanetcenter.net}
     {https://minorplanetcenter.net}} revealed that no known small bodies could have been found projected towards the regions
     surveyed at the observed times. The high ecliptic latitude of the imaged fields precluded any detection of new members of the 
     main asteroid belt and new members of the near-Earth object (NEO) populations were not expected to be detected due to the 
     slow cadence of our survey. In addition to the targeted {\it TESS} candidates, our survey may only have detected relatively 
     slow moving objects with high orbital inclination. No such objects emerged during the data analysis. 
     \citet{2000mbos.work...87F} pointed out that, when observing towards the ecliptic and a few tens of degrees from it, the 
     apparent sky density of regular TNOs translates into one object brighter than $R$=23.0~mag per square degree of sky. 
     Therefore, not finding any TNOs in over one square degree of sky at high ecliptic latitude is consistent with the 
     expectations. The fields studied are seldom observed by surveys targeting the outer Solar system.

  \section{Discussion}
     As pointed out above, we failed to recover candidates~9 and 11 identified by \citet{2020PSJ.....1...81R}. In getting a null
     result, we must provide some indication of the sensitivity of our study. In a survey such as this one, the typical issues 
     that would negatively impact sensitivity are: having an improper cadence, incorrect survey area size (perhaps too small), or 
     insufficient depth. Here, we discuss these three sources of lack of sensitivity and their possible effects on our results. As 
     shown in Table~\ref{observations}, the Moon could not possibly have been a factor in the non-detection of the candidates as 
     the sky was dark during the time intervals used on-sky.

     In our study, each field in the survey area was repeatedly revisited at variable cadence intervals. Each main field (one in
     Cepheus and one in Camelopardalis) was sampled with 77 pointings that took about one hour to be completed.
     Table~\ref{observations} shows that the seeing was variable during the observations; however, the range of variability was
     0{\farcs}6--1{\farcs}3 with the exception of the last night of the survey, July 8, when the spanning interval was 
     (1{\arcsec},~2{\arcsec}). In general and for each night and field, the quality of the images associated with the 77 pointings 
     is reasonably homogeneous. Then, Table~\ref{observations} shows that each main field was revisited with a cadence of 24~h 
     (first three nights) or 48~h (last two nights). The rate of motion of a Solar system body located at a heliocentric distance  
     $s$ is maximum at opposition and minimum near quadrature. The actual values can be estimated under the circular and coplanar 
     orbits approximation using (see e.g. \citealt{1999AREPS..27..287J,2000mbos.work...87F}):
     \begin{equation}
        \mu_{\rm opp} = \frac{3547.2}{s + \sqrt{s}}\ \ \ {\rm arcsec}\ {\rm d}^{-1}\,, \label{oppo}
     \end{equation}
     \begin{equation}
        \mu_{\rm qua} = \frac{3547.2}{s^{3/2}}\ \ \ {\rm arcsec}\ {\rm d}^{-1}\,, \label{quad}
     \end{equation}
     where 3547.2 is the mean orbital angular motion of our planet in arcseconds per day and $s$ is measured in au. These 
     approximations are good enough for objects observed near perihelion or aphelion (as it could be the case for some {\it TESS} 
     candidates); in a more general situation, the rate of motion considering a highly eccentric orbit might be a factor two 
     slower than in the circular approximation. The impact of the inclination is rather negligible. For candidate~9, 
     $s$$\sim$200~au so $\mu_{\rm opp}$=16{\farcs}6~d$^{-1}$ and $\mu_{\rm qua}$=1{\farcs}3~d$^{-1}$; for candidate~11 with 
     $s$$\sim$160~au, the respective values are 20{\farcs}5~d$^{-1}$ and 1{\farcs}8~d$^{-1}$. As the seeing was mostly 
     $<$1{\farcs}3 and the cadence was $\geq$24~h, it is virtually impossible that the cadence intervals may have played any role 
     in the non-detection of the candidates.

     Our survey scanned square patches of sky of over 1{\degr}$\times$1{\degr} (see Fig.~\ref{fields}) that may perhaps be not
     large enough to ensure that the candidates appear projected towards the region imaged. The only relevant data from
     \citet{2020PSJ.....1...81R} are the values of the equatorial coordinates of the objects (referred to epoch 2458838.92~JD or
     10:4:48.00 UT on December~21, 2019 and 2458810.25~JD or 18:0:0.00 UT on November~22, 2019) and the geocentric distance (in
     the range 80--200~au). It is important to realize that due to parallax, the apparent position of these candidates will trace
     out an ellipse on the sky over the course of a year \citep{2005ApJ...635..711G}. Such a parallactic ellipse will have a
     maximum dimension of 17{\farcm}2 for a candidate located 200~au from the Sun and 21{\farcm}5 for the one at 160~au; the 
     ellipse will have maximum eccentricity for those objects with very low values of the ecliptic latitude and it will be nearly 
     circular for those close to the ecliptic pole. In addition to the apparent parallactic motion, the candidates will have a 
     proper motion due to its intrinsic displacement as they go around the barycentre of the Solar system. When observing at 
     quadrature, Earth's parallactic motion is close to zero; therefore, the observed rate of motion is the direct result of the 
     object's orbital motion and can be estimated using Eq. (\ref{quad}). For a range in distance of 80--200~au, the proper motion 
     values are in the range 30{\farcm}2 to 7{\farcm}6 per year, respectively, with 10{\farcm}7 per year for a distance of 160~au. 
     Unless the true values of the relevant parameters of the candidates deviate significantly from those in table~2 of
     \citet{2020PSJ.....1...81R}, the survey area size should have been sufficient to enable detections, particularly in the case
     of the most distant candidate.

\begin{figure}
       \centering
        \includegraphics[width=\linewidth]{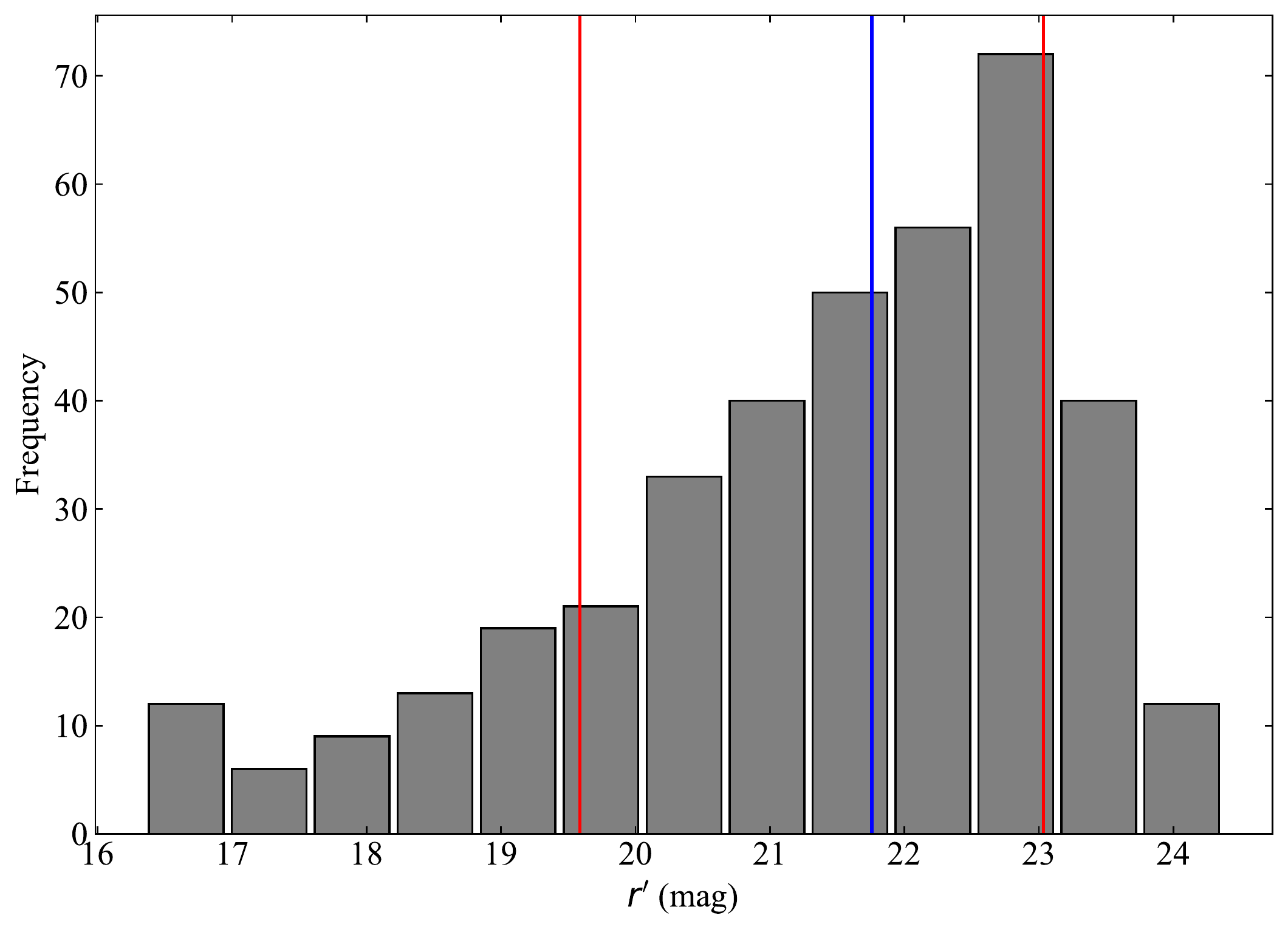}
        \includegraphics[width=\linewidth]{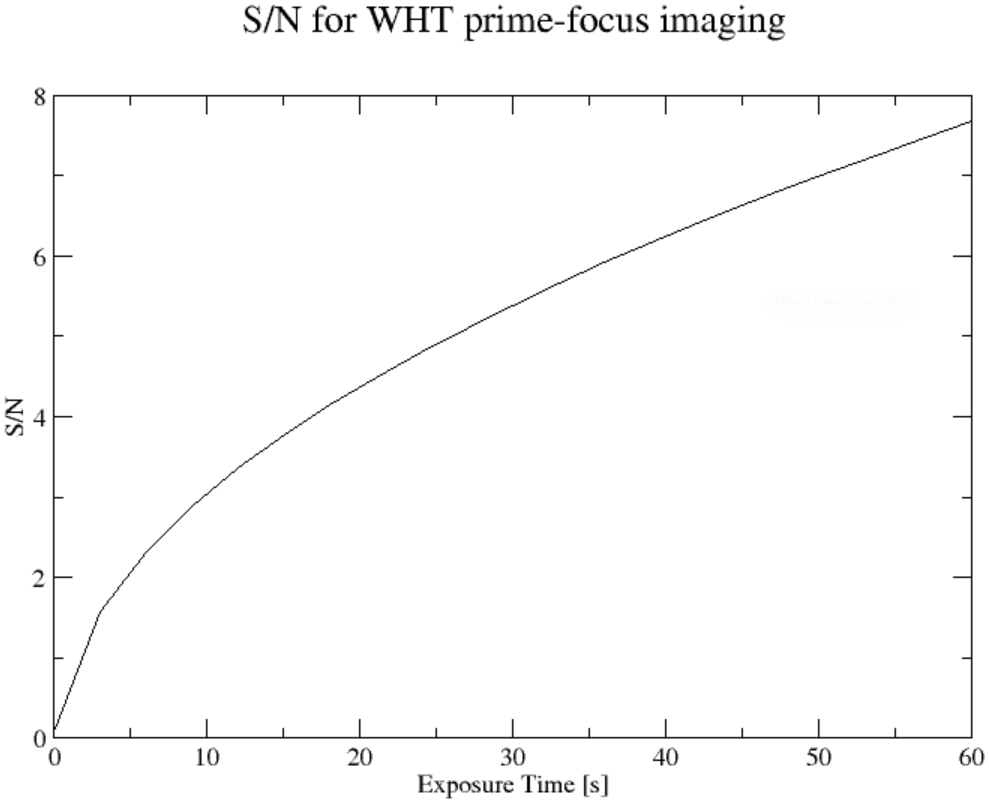}
        \caption{Completeness limit and S/N. The top panel shows the magnitude distribution of the stars detected from a 
                 representative pointing in Cepheus, observed at an elevation of 25{\degr} (using Pan-STARRS DR1 as a reference 
                 catalog, \citealt{2020ApJS..251....7F}). The median is shown in blue and the 16th and 84th percentiles are in 
                 red. In the histogram, bins were computed using the Freedman and Diaconis rule \citep{FD81}. The bottom panel 
                 displays the S/N for {\it WHT} prime-focus imaging as a function of the exposure time from the ETC of the Red+4 
                 instrument with a seeing of 0{\farcs}9 at an airmass of 2.3 and $R$=23.2~mag. A S/N ratio close to 5.5 is 
                 expected to be reached for a 30~s integration.
                }
        \label{lim}
     \end{figure}

     The fields targeted by our proof-of-concept mini-survey were observed at relatively high airmasses (see Section~2). 
     Therefore, it can be argued that our null result could be due to enhanced extinction caused by observing at such low 
     elevations. However, this is likely not the case because the same EURONEAR collaboration carried out a nearly concurrent 
     mini-survey looking for Atira and Vatira asteroids (see e.g. \citealt{2012Icar..217..355G}) using the same instrumental setup 
     and similar automated reduction pipeline. Atiras or Interior Earth Objects (IEOs) have aphelion distances $<$0.983~au and can 
     only be observed at low solar elongations (often $<$70{\degr}); Vatiras have aphelia $<$0.718~au, are observed at very low 
     solar elongations (typically $<$40{\degr}--45{\degr}), and just one such object is currently known 
     \citep{2020MPEC....A...99B,2020MNRAS.494L...6D,2020MNRAS.493L.129G}, 594913 `Ayloi'chaxnim (2020~AV$_{2}$). Although the 
     nearly concurrent EURONEAR small-scale survey (12 nights with an average time of 1.5~h per night devoted to searching 
     for Vatira and Atira asteroids towards the evening and early morning skies) did not find any new Vatiras or IEOs, it did 
     observe at elevations in the range 15{\degr}--30{\degr} and recovered known objects at apparent magnitudes close to or above 
     23 --- for example 2017~HO$_{20}$, a main belt asteroid that was observed near conjunction. Furthermore, Fig.~\ref{lim}, 
     top panel, shows the $r'$-magnitude distribution of the detected stars from one representative pointing out of the 77 
     ($\times$5) pointings intended to recover the {\it TESS} candidate in Cepheus. The histogram was produced using the 
     \textsc{Matplotlib} library \citep{2007CSE.....9...90H} with sets of bins computed using \textsc{NumPy} 
     \citep{2011CSE....13b..22V,2020Natur.585..357H} by applying the Freedman and Diaconis rule \citep{FD81}. Figure~\ref{lim}, 
     bottom panel, shows the expected S/N for {\it WHT} prime-focus imaging as a function of the exposure time from the exposure 
     time calculator (ETC)\footnote{\href{http://catserver.ing.iac.es/signal/index.php}
     {http://catserver.ing.iac.es/signal/index.php}} for a similar camera (Red+4, PFIP) for observations carried out under a 
     seeing of 0{\farcs}9 at an airmass of 2.3 in dark time and $R$=23.2~mag. TNOs have color index $V-R$ in the range 0.2--1.2 
     (see e.g. \citealt{2015A&A...577A..35P}). Consistently, we are confident that we reached a limiting magnitude $r'$=23.0~mag, 
     which actually corresponds to $V$$>$23.0~mag (in fact, it could be 23.2--24.2~mag). Therefore, our distant TNO candidate 
     survey could have detected candidates~9 and 11 in table~2 of \citet{2020PSJ.....1...81R} unless both candidates are much 
     fainter than predicted or they are false positives in {\it TESS} data. In any case, \citet{2020PSJ.....1...81R} stated that 
     many if not most of the high signal significances reported in their table~2 could be the result of unmodelled systematic 
     errors. On the other hand, in the fields observed towards Cepheus, there are a few bright stars that might have hidden a 
     putative moving object with the properties of candidate~9; we estimate that the probability of having missed the candidate as 
     a result of it moving projected towards one of those bright stars is $<$1 per cent (considering the area affected by the 
     bright stars and associated diffraction spikes).
%
%
%
%

  \section{Summary and conclusions}
     We have performed, to our knowledge, the first ever search to recover outer Solar system body candidates found in data from 
     NASA's {\it TESS} mission. We surveyed, in Sloan $r'$, two areas of sky (square patches of over 1{\degr}$\times$1{\degr}) 
     around the circumpolar {\it TESS} candidates located at (07$^{\rm h}$:00$^{\rm m}$:15$^{\rm s}$, 
     +86{\degr}:55{\arcmin}:19{\arcsec}), 202.8~au from Earth (candidate~9 in \citealt{2020PSJ.....1...81R}), and 
     (06$^{\rm h}$:39$^{\rm m}$:47$^{\rm s}$, +83{\degr}:43{\arcmin}:54{\arcsec}) at 162.1~au (candidate~11), with the 4.2-m 
     {\it WHT} and a QHY600L CMOS camera mounted on-axis at its prime focus. Our results are summarized as follows.
     \begin{enumerate}[(i)]
        \item We were unable to recover either of the candidates at Sloan $r'$$\leq$23~mag.
        \item After considering both visibility and detectability issues we interpret our null detections as evidence that either
              both candidates are actually much fainter than predicted or that they are false positives in {\it TESS} data.
     \end{enumerate}
     These findings emphasize the need for independent confirmation of detections of distant Solar system object candidates 
     resulting from blind shift-and-stack searches.  

  \section*{Acknowledgements}
     We thank the referee for a constructive and useful report. This work was developed within the framework of the EURONEAR 
     collaboration and it is based on observations made with the {\it William Herschel} Telescope (operated by the {\it Isaac 
     Newton} Group of Telescopes, ING), proposals SW2021a24 and SW2021a10. We are thankful to the following ING staff who planned, 
     carried out, and reported our service observations: Lara Monteagudo, Norberto Gonzalez, Yeisson Osorio, and Richard Ashley. 
     CdlFM and RdlFM thank A.~I. G\'omez de Castro for providing access to computing facilities, and S. Deen and J. Higley for 
     comments on TNOs. This work was partially supported by the Spanish `Ministerio de Econom\'{\i}a y Competitividad' (MINECO) 
     under grant ESP2017-87813-R and the `Agencia Estatal de Investigaci\'on (Ministerio de Ciencia e Innovaci\'on)' under grant 
     PID2020-116726RB-I00 /AEI/10.13039/501100011033. In preparation of this Letter, we made use of the NASA Astrophysics Data 
     System and the MPC data server. This research has made use of SAOImage DS9, developed by the Smithsonian Astrophysical 
     Observatory.

  \section*{Data Availability}
     The data underlying this article will be shared on reasonable request to the corresponding author.

  \bsp
  \label{lastpage}
\end{document}